\documentclass[english,aps,pra,twocolumn,reprint,superscriptaddress]{revtex4-1}
\usepackage[T1]{fontenc}
\usepackage[utf8]{inputenc}
\usepackage{color}
\usepackage{amstext}
\usepackage{graphicx}

\makeatletter

\newcommand{\lyxdot}{.}

\makeatother

\usepackage{babel}
\begin{document}

\preprint{This line only printed with preprint option}

\title{Dynamic process of free space excitation of asymmetry resonant microcavity}

\author{Fang-Jie Shu}

\email{shufangjie@gmail.com}

\affiliation{Department of Physics, Shangqiu Normal University, Shangqiu, He'nan
476000, P. R. China.}

\author{Chang-Ling Zou}

\affiliation{Key Laboratory of Quantum Information, University of Science and
Technology of China, Hefei, Anhui 230026, P. R. China}

\author{Fang-Wen Sun}

\email{fwsun@ustc.edu.cn}

\affiliation{Key Laboratory of Quantum Information, University of Science and
Technology of China, Hefei, Anhui 230026, P. R. China}

\date{\today}
\begin{abstract}
The underlying physics and detailed dynamical processes of the free
space beam excitation to the asymmetry resonant microcavity are studied
numerically. Taking the well-studied quadrupole deformed microcavity
as an example, we use a Gaussian beam to excite the high-Q mode. The
simulation provides a powerful platform to study the underlying physics.
The transmission spectrum and intracavity energy can be obtained directly.
Irregular transmission spectrum was observed, showing asymmetric Fano-type
lineshapes which could be attributed to interference between the different
light paths. Then excitation efficiencies about the aim distance of
the incident Gaussian beam and the rotation angle of the cavity were
studied, showing great consistence with the reversal of emission efficiencies.
By projecting the position dependent excitation efficiency to the
phase space, the correspondence between the excitation and emission
was demonstrated. In addition, we compared the Husimi distributions
of the excitation processes and provided more direct evidences of
the dynamical tunneling process in the excitation process.
\end{abstract}
\maketitle

\section{introduction}

Optical microcavity has drawn many attentions owe to its high intracavity
field intensity \cite{vahala2003}. Light-matter interactions, such
as laser \cite{qian1986lasing}, nonlinear optics \cite{matsko2005review},
cavity quantum electrodynamics \cite{haroche1989cavity,xiao2012strongly},
sensors \cite{vollmer2008whispering} and optomechanics \cite{park2009resolved},
can benefit from the microcavities. No matter these applications are
based on active or passive microcavities, energy exchange with the
outside is essential. Variously shaped asymmetry resonant microcavities
(ARCs) (quadrupole \cite{nockel1997ray,gmachl1998}, spiral \cite{chern2003unidirectional},
stadium \cite{fang2005analysis}, Limacon \cite{wiersig2008combining},
and half-quadrupole-half-circle \cite{xiao2009low,zou2009ArXiv})
were extensively studied recently since light can be collected more
efficiently and conveniently through free space in those cavities than in symmetry circular
or spherical cavity.

However, most studies put emphasis on the directional emission for
efficient collection \cite{xiao2012strongly,gmachl1998,chern2003unidirectional,fang2005analysis,wiersig2008combining,zou2009ArXiv,xiao2009low,lee2011PRA},
but the inverse processes, i.e. the free space excitation \cite{park2009resolved},
have not been fully explored. An’s group performed quadruple droplet
pumping experiments, then put mode-coupling theory and numerical
simulation forward to explain the experimental phenomenon \cite{lee2007APL,yang2008APL,yang2010OE}.
Their studies focused on the laser emission in a full-chaotic cavity
by a plane wave excitation, and the process of chaotic transport in
the pumping process was studied indirectly from the laser threshold.
\textcolor{black}{However, threshold also depends on the overlap between
the pumping mode and lasering mode, and there is also mode competition
in the multimode cavity. The laser threshold may not linearly depend
on the excitation efficiency of pumping mode.} Therefore, the direct
evidence of the efficient pump to a passive high-Q modes, and the
studies of underlying principle are required. In addition, more perspectives
of the free space coupling, such as the focused beam pump and the
spectrum, are lacking.

In this paper, the free space excitation of an ARC by a focused Gaussian
beam is studied numerically by boundary elements method (BEM). Then
the field distribution, mean intracavity intensity (MII) and the transmission
spectrum varying with frequency of the incident beam are also studied.
Fano-type lineshape of spectrum is investigated from interference
point of view. Excitation efficiency represented by MII is studied
by adjusting relative position of the incident beam and the cavity.
A comparison between excitation and emission is performed with the
help of ray optics. Finally, resorting the Husimi projection of the
wave field and the projection of the MII distribution, we discuss
the physics mechanism of the free space excitation in several cavities
with different degree of deformations.

\section{the basic dynamics of quadrupole shaped microcavity}

\begin{figure}[h]
\begin{centering}
\includegraphics[width=8.5cm]{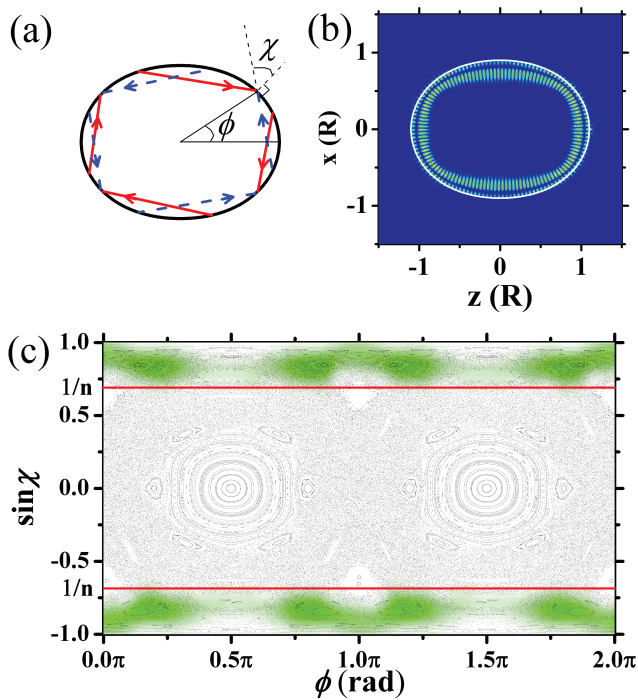}
\par\end{centering}

\caption{(a) Period-4 rectangle obits. The ray trajectories are calculated
from the location of maximal points in the Husimi projection in (c).
Considering the Goos-Hänchen shift, clockwise (red solid line) and
counter clockwise (blue dash line) obits split with each other, and
the obits are not closed. (b) Near field intensity of the order-2
mode. (c) Poincaré Surface of Section (SOS) of the quadrupole cavity
(black dots) and Husimi projection (green areas) of the WGM illustrated
in (b). In the SOS, ray is recorded as a point with coordinate ($\phi$,
sin$\chi$), where $\phi$ is the polar angle of incident point, and
$\chi$ is incident angle on the boundary, respectively. The whole
SOS of a quadrupole cavity is mixed with Kolmogorov–Arnold–Moser (KAM)
torus (curves near the upper and lower boundary of SOS), islands (close
curves), and chaotic sea (collection of other dispersed points) structure.\label{fig.sos}}
\end{figure}

Here, we focus on the two dimensional (2D) quadrupole microcavity,
which has been studied extensively in theory and experiments \cite{schafer2006directed}.
Basic features and underlying physics found in this specific cavity
can also extend to other ARCs, like the half-quadrupole-half-circle
microcavity. The boundary of quadrupole microcavity in polar coordinate
is described by $r(\phi)=R(1+\varepsilon\cos2\phi)$ with radius\textit{
R} and degree of deformation $\varepsilon$ \cite{nockel1997ray}.
In the deformed cavity, the ray dynamics shows quasi-periodic, periodic
and chaotic trajectories, corresponding to the regular KAM torus,
islands, and chaotic sea in the phase space presentation. Fig. \ref{fig.sos}(c) (black dots)
shows the ray dynamics in quadrupole cavity with deformation $\varepsilon$ =
0.1 as an example. Energy of a
high quality factor mode is always stored in the vicinity of the regular
structure above the critical line by total internal reflection [Fig.
\ref{fig.sos}(a)], part of energy tunnels to the nearby chaotic zone
and finally leaking out the cavity by reflections. The reflections
condense in a few trails, which are called manifold, in phase space
[Fig. 5(a)]\cite{schwefel2004dramatic}. The asymmetrical nature of
the manifold results in the directional emission.

Ray model just gives basic physics insights of the ARCs. For more
underlying physics and precise details, we should also consider the
aspect of wave nature of light. The wave functions of high-Q WGMs
are calculated numerically (solving Maxwell function using boundary
elements method (BEM) \cite{wiersig2003BEM}). Setting uniform cavity
refractive index \textit{n} = 1.45, we search the resonances with
quick root searching method \cite{zou2011OE} at $kR\sim50$, where
\textit{kR} is also the dimensionless cavity size scale to beam wavelength.
Figure \ref{fig.sos}(b) is a eigenmode of quadrupole cavity near
\textit{kR} = 50, it is known as the order-2 mode for existing two
intensity peaks along the radial direction. We can see the ray-wave
correspondence by Husimi projection of the wave function in phase
space, as is shown by the green contour in Fig. 1(c). Eight obviously
condensed regions are clockwise and counter-clockwise period-4 islands,
which corresponds to the period-4 orbits in Fig. 1(a). Noting that
the orbits shows a displacement at each reflection, which is attributed
to the Goos-Hänchen shift \cite{goos-hanchen2008}. In the following
studies, we focus on this mode to study free space excitation.

\section{The configuration of Gaussian beam excitation}

\begin{figure}
\centering{}\includegraphics[width=5cm]{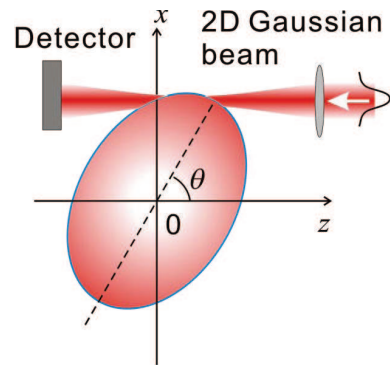}\caption{A schematic diagram of an incidence beam, a cavity, and a detector
used in the simulation. Note that this configuration is conjugated
with former experiments as aim distance $x=-x_{exp}$ and rotation
angle $\theta=\frac{\pi}{2}-\theta_{exp}$ \cite{yang2008APL}. \label{fig.config}}
\end{figure}

Now, we turn to study the free space coupling to the WGM {[}Fig. 1(b){]}
by Gaussian beams {[}Fig. \ref{fig.config}{]}. From the perspective
of wave optics, incident focused 2D Gaussian beam reads \cite{oron2004third-harmonic}
\begin{equation}
A(x,z)=\frac{1}{\sqrt{1-i\zeta}}exp[-\frac{(x-x_{0})^{2}}{\omega_{0}^{2}(1-i\zeta)}-ikz].\label{eq:gaussian}
\end{equation}
Here $\zeta=2z/b$ is the dimensionless longitudinal location defined
in terms of the confocal length $b=k\omega_{0}^{2}$, and $\omega_{0}$
is the beam waist. This Gaussian beam is set as $\omega_{0}=0.3R$,
and focus on \textit{z} = 0 line. The relatively narrow waist of the
Gaussian beam can be obtained in an experiment by a convergent lens,
and benefit the coupling efficiency.

The coupling efficiency depends on the excitation condition of the
Gaussian beam, i.e. the direction of the beam and the excitation point
on the cavity boundary. So, we rotate the cavity around the original
point, and shift the incident beam parallel along \textit{x} axis.
Thus, the coupling config can be represented by two parameters $(\theta,x)$,
where $\theta$ is angle from axis z to major axis of the quadrupole,
and $x$ is the aiming distant between the axial line of the Gaussian
beam and the central point of the cavity, which is just the numerical
value of \textit{x} coordinate of the axial line in our coordinates
system.

\begin{figure}
\begin{centering}
\includegraphics[width=8.5cm]{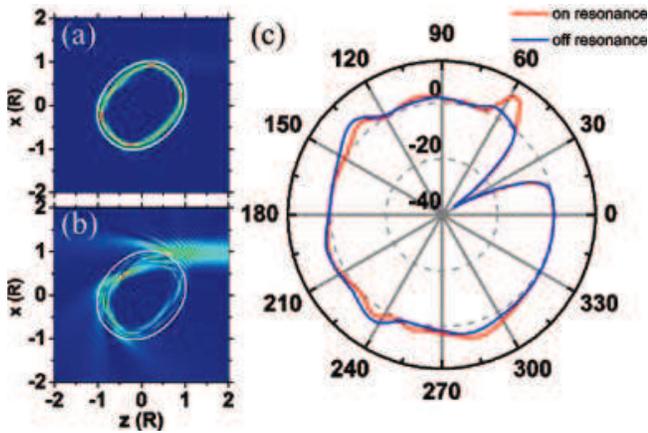}
\par\end{centering}

\caption{False color intensity patterns of excited cavity using on-resonance
(a) and off-resonance (b) incident beam. White lines are the boundaries
of the cavity. (c) Angular distributions of outward energy flow through
a circle with radius 1.8\textit{R} correspond to the pattern (a) (red
line) and (b) (blue line), respectively. A positive value indicates
the outflow, a negative value indicates inflow. Parameters: \textit{kR}
= 49.726 (on), 49.9 (off), $\phi=45^{o}$, \textit{x} = 0.97, $\varepsilon$
= 0.1. \label{fig.onoffpattern}}
\end{figure}

In the first place, an on-resonance and an off-resonance excited patterns
of light intensity in cavity with $\varepsilon=0.1$ are shown in
Figs. \ref{fig.onoffpattern}(a) and \ref{fig.onoffpattern}(b). All
wave simulations are performed by BEM, which solves the Helmholtz
equation of the electromagnetic field with particular boundary conditions.
For simplicity but without loss of generality, only transverse magnetic
mode is taken into account. In this polarization, electric field vector
is vertical to the cavity plane. The field amplitude and its derivative
are both continuous on the interfaces of dielectrics. Incident beam
frequency \textit{kR} = 49.726 (49.9) is chosen for on-resonant (off-resonant)
coupling, $\theta=45^{\circ}$, and $x=0.97$ are set for getting
high intracavity intensity. As is shown in Fig. \ref{fig.onoffpattern},
normalized 2D Gaussian incident beam is partly coupled into the cavity.

In the on-resonance case, notable intensity lies near the cavity boundary
forming the whispering gallery like mode, which is consistent with
a traveling wave mode corresponding to the order-2 resonance mode
at the same \textit{kR} {[}Fig. \ref{fig.sos}(b){]}. There is no
interference\textcolor{red}{{} }\textcolor{black}{fringe} along the
$\phi$ in Fig. 3(a), which is due to only counter-clockwise traveling
WGM excited in such excitation configure with a focus Gaussian beam
pumping away from the center of the cavity. When it comes to off-resonance
case, the cavity field is very different from the eigenmode field.
The incidence beam will be refracted into the cavity,  reflect along
the cavity boundary several times, and finally leak out by refraction\textcolor{red}{{}
}\textcolor{black}{\cite{hentschel2001}}. Due to the chaotic ray
dynamics, the rays show large expansion in the $r$ direction.\textcolor{red}{{}
}\textcolor{black}{This large expansion can also be interpreted as
excitation of multiple low-Q modes. The field of which shows higher
radial quantum number. These multiple interference also occurs in
the on-resonance case, but is not visible due to high resonant field
intensity.}\textcolor{red}{{} }\textcolor{black}{Note that the incident
Gaussian beam is also not visible in the on-resonance condition, which
is due to the greatly enhanced cavity field making the relatively
low intensity of pumping field invisible.}\textcolor{red}{{} }In order
to evaluate the intensity in the cavity, we introduce the MII as average
intensity index. When the maximum intensity of the incident Gaussian
beam on the waist line is set to a unit as reference intensity, the
MII is proportional to coupling efficiency and the Q of a mode. For
a particular mode, such as order-2 mode that we concerned, Q is fixed,
so the MII represents the coupling efficiency directly. The MIIs are
5.02 and 0.23 for on-resonance and off-resonance excitation, respectively.
MII of on-resonance excitation is much larger than that of off-resonance
excitation. So in the figure of relative intensity pattern {[}Fig.
\ref{fig.onoffpattern}(a){]} Gaussian beam leaves only a dim trace
against to the strong intracavity intensity.

Moreover, angular distributions of radial energy flows of on-resonance
and off-resonance {[}Fig. \ref{fig.onoffpattern}(c){]} were calculated
by the Poynting vector in radial direction\textcolor{black}{{} $\bar{\vec{p}}\cdot\vec{n}=\textrm{Im}(E*\cdot\partial_{v}E)/2\omega\mu$}
\cite{zou2009accurately} on a circle with radius of 1.8\textit{R}
and center on the original point, where \textcolor{black}{$\partial_{v}=\vec{v}\cdot\nabla$
}is the normal derivative on radial direction \textcolor{black}{$\vec{n}$,
$\omega$ is angular frequency, and $\mu$ is }relative permeability.
The flow is negative at 30 degree, which shows the position of the
incident Gaussian beam, while the other region is always positive
which indicates the outgoing field. For both on- and off- resonance,
the net flow as the sum of energy flow in all direction is zero, which
indicates that the energy conservation is preserved, i.e. the ideal
passive cavity (with real refractive index) does not produce or consume
energy. It can be seen from the Fig. \ref{fig.onoffpattern}(b) that
the energy flow distribution of off-resonance excitation mainly embodies
the redistribution of the incident energy by cavity-caused scattering.
In the on-resonance case, there is also the emission from resonant
modes. So, two emission peaks of the order-2 CCW mode emerge near
$60^{\circ}$ and $240^{\circ}$ on detector circle in Fig. \ref{spectrum}(c),
respectively. The interference between the emission of high-Q mode
and reflection light of multiple beam also give rise to the modulation
of distribution of energy flow in other directions.

\section{The transmission spectrum}

\begin{figure*}
\begin{centering}
\includegraphics[width=13cm]{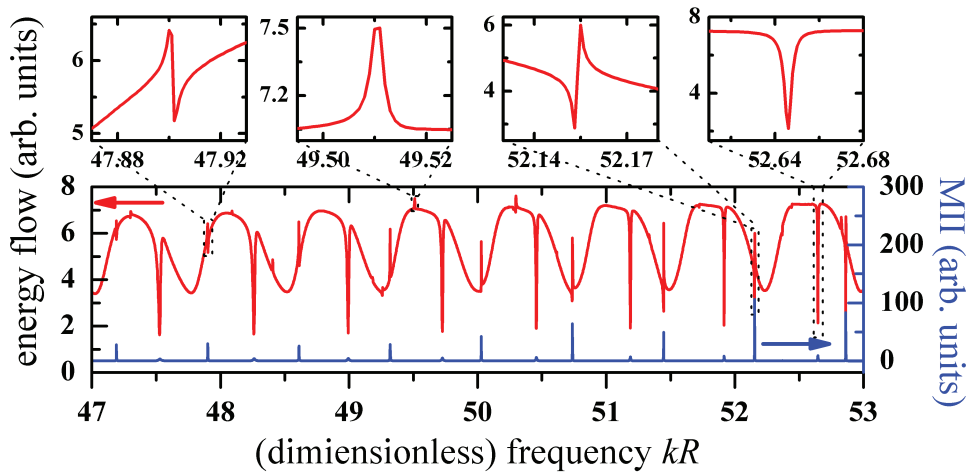}
\par\end{centering}

\caption{(color online) Spectrum of energy flow (red curves) and mean intracavity
intensity (blue curves). The enlarged scales are four typical lineshapes
in transmission spectrum.}
\label{spectrum}
\end{figure*}

Similar with experiment, through a frequency scan to the incident
beam, we can obtain the spectrum by using a detector with fixed relative
positions of incident Gaussian beam and microcavity {[}with $\theta=45^{\circ}$
and $x=0.97${]}. The energy {[}Fig. \ref{spectrum} red curves{]}
is the integration of the outward energy flow through the line detector.
In addition, the MII spectrum was also recorded {[}Fig. \ref{spectrum}
blue curve{]}. The transmission spectrum near $kR=50$ shows the shape
of the periodic low-frequency harmonized wave superimposed with high-frequency
spikes. Therein, the low-frequency harmonized modulation is interference
of light that undergo different quasi-period rounds \cite{hentschel2001}.
\textcolor{black}{The modulation period is about 0.7 ~ 1/n,
i.e. roughly corresponding to an optical distance of one round trip.}

On the periodically modulated spectra background, there are many sharp
symmetry dips, peaks and asymmetric Fano-type line shapes in the transmission
spectrum. Comparing with the MII spectra, these sharp variations correspond
to a magnified spikes in MII. The correspondence indicates that the
formation of those line shapes is related to the on-resonance excitation.
We note that the MII spectrum clearly shows two sets of resonance
peaks. The set with larger intervals of adjacent lines and larger
MII is order-1 modes and the other is order-2 modes which are concerned
in this paper. Different from the Lorentz line shape in the transmission
spectrum of waveguide coupled cavity, here the line shapes are formed
by the interference between emission light of resonance CCW mode and
the scattered light by multiple beam interference.

There are two different paths for light to transmit from source to
detector: (1) the light coupling to the high-Q cavity mode and emitted
from the mode, i.e. direct scattering from high-Q mode (2) the light
transmitted after multiple beam interference, or we can say scattering
from multiple low-Q modes. The final collected energy is given by

\begin{equation}
E(\delta)=E_{mul}e^{i\alpha}-E_{res}e^{i\beta}\frac{1}{i\delta/\kappa+1}.\label{eq:gaussian-1}
\end{equation}
Here, $\delta$ is the frequency detuning, $\kappa$ is the linewidth
of high-Q mode, $\alpha$ and $\beta$ are phase factors that denotes
the phase variance between source and detector, $E_{mul}$ and $E_{res}$
are the amplitude (real number) of the multiple beam interference
and resonance scattered light. For the multiple beam interference,
the low-Q modes involved have linewidth much larger than $\kappa$$,$
thus the phase $\alpha$ is almost constant for $\delta/\kappa\sim1$.
From the formula, the phase difference $\Delta=\alpha-\beta$ determines
collected transmission spectra. (i) $\Delta=0$ gives symmetry Lorentz
lineshape dip (the 4th line of the upper panel in Fig. \ref{spectrum}).
(ii) $0<\Delta<\pi$ is left-low-right-high asymmetry Fano shape (the
3rd line of the upper panel in Fig. \ref{spectrum}). (iii) $\Delta\sim\pi$
gives symmetry Lorentz lineshape peak (the 2nd line of the upper panel
in Fig. \ref{spectrum}). (iv) $\pi<\Delta<2\pi$ shows left-high-right-low
asymmetry Fano shape (the 1st line of the upper panel in Fig. \ref{spectrum}).
The $\Delta$ varies with the frequency of the incident light, thus
we observed all four different lineshape types in the transmission
spectra. Then, interference line shapes are formed. But unlike the
waveguide coupled case, the energy flow irradiates into all directions
in free space. So the spectrum collected by a detector in a particular
direction has no energy conservation relationship with MII, which
makes the evaluation of pumping efficiency from transmission spectrum
invalid. That is why we turn to choose the MII as pumping efficiency
indicator. In addition, the phase also varies according to different
position of the detector making different lineshapes\textcolor{black}{{}
\cite{zou2012theory}}.

\section{Details of excitation process}

\begin{figure}
\begin{centering}
\includegraphics[width=8.5cm]{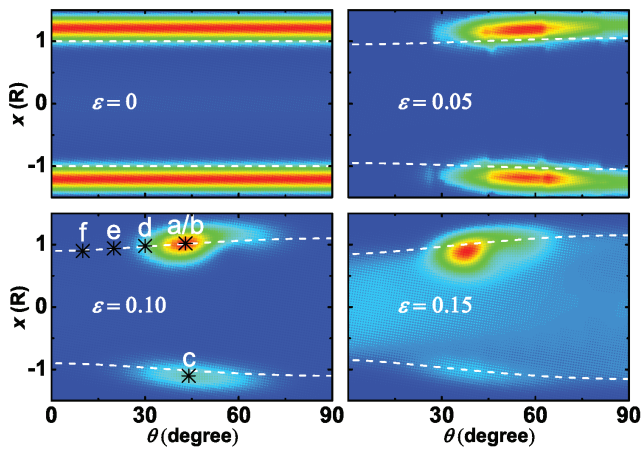}
\par\end{centering}

\caption{Pump efficiency distributions of quadrupole cavities with $\varepsilon=0.00,0.05,0.10,0.15$
in real space position coordinates $(\theta,x)$. The frequency of
incident beam fixed on resonance with the \textit{q} = 2 mode. White
dash lines are the projection boundary of cavity in \textit{x} axis.
Because of the decrete symmetry, it is sufficient only $0^{\circ}<\theta<90^{\circ}$
is investigated.\label{fig:Pumpefficiency}}
\end{figure}

The resonant mode of deformed quadrupole microcavity has its favorite
emission direction. Hence the free space coupling should be the most
effective under a particular pumping condition, as a reversal of directional
emission from deformed microcavity\textcolor{black}{{} \cite{zou2012theory}}.
Here, we studied the pumping with different parameters ($\theta$,$x$)
{[}Fig. 2{]}, and recorded the MIIs of four quadrupole microcavities
($\epsilon=0.00,0.05,0.10,0.15$) as illustrated in Fig. \ref{fig:Pumpefficiency}.
MIIs of circular cavity ($\epsilon=0$) do not change with the rotation
angles because of the rotational symmetry of the cavity. Moreover,
the maximum of MII occurs beyond the boundary, which indicates the
most effective way of excitation the circular cavity is barrier tunneling.
The slightly deformed quadrupole cavity with deformation of 0.05 has
angle selectivity. Nevertheless, the main excitation manner is still
tunneling, since the maximum pump position is \textcolor{black}{outside
t}he cavity too. In addition, MII distribution of this cavity shows
a roughly horizontal symmetry, which is a result of the tunneling
coupling showing CW and CCW symmetry. However, in quadrupole cavity
with deformation of 0.10, the highest MII area is located near the
boundary projection line, i.e. incident beam is tangential to the
boundary of the cavity. Because the width of the incident beam is
not zero, half of the lights enter into the cavity by refraction.
The size of the strong excitation region in the $\theta-x$ figure
shrinks significantly comparing with the slightly deformed case. This
localization suggests strengthening of selectivity of pumping angle.
In addition, the MII of highest excitation region in the upper part
of the figure is much stronger than that of the second highest excitation
region lying in the lower part. \textcolor{black}{Different from the
barrier tunneling coupling where the tunneling rate only depends on
the local curvature and is same for CW and CCW wave, the dynamical
tunneling coupling is sensitive to the energy distribution of the
incident beam in the chaotic sea of phase space. The dynamics of rays
incident in these two regions are non-identical due to the asymmetric
unstable manifold in the chaotic sea, thus the energy distributions
in cavity for these two excitation conditions are different, and finally
raises different excitation efficiency to high-Q modes by dynamical
tunneling.} By increasing the deformation to 0.15, the highest MII
zone is mainly located inside the boundary projection line, which
suggests refraction excitation. This is the only optimal position
as the dynamical tunneling dominats the pumping process. And the MII
is visible when a beam shines on the cavity which is due to low contrast
to the optimal coupling. The MII is always nonzero when light shines
on the cavity.

\begin{figure}
\centering{}\includegraphics[width=8.5cm]{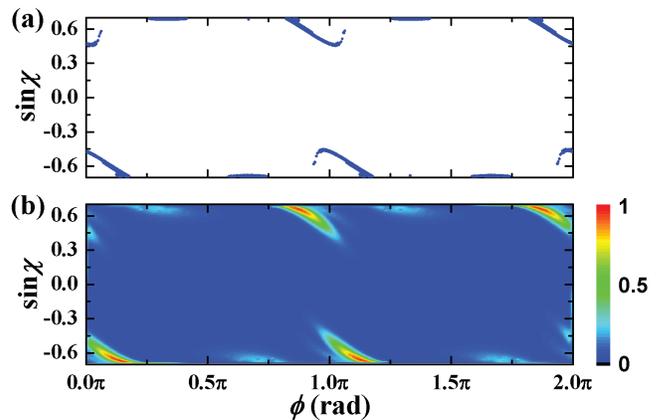}\caption{(a) Emissions of the quadrupole cavity with $\varepsilon=0.1$ predicted
with ray optics. The points in the figure are rays whose reflection
angles are larger than the critical total reflection angle. So, collections
of refractions of those rays constitute the emission picture. (b)
Phase space presentation of $\varepsilon=0.1$ panel of Fig. \ref{fig:Pumpefficiency}.\label{fig:projection}}
\end{figure}

As the inverse process of emission, excitation of an ARC should show
some correspondence with directional emission. For example, comparing
our pumping efficiency with the emission figure obtained in Ref. \cite{lee2011PRA},
they show great consistence. In order to express this correlation
directly, MII distribution ($\varepsilon=0.1$ panel of Fig. \ref{fig:Pumpefficiency})
is projected to phase space {[}Fig. \ref{fig:projection}(b){]} by
ray trajectory. First, each point in real space distribution is treated
as rays. The rays impacted on the cavity (points between the white
dash line) can refract into the cavity. Then, record pairs of refracting
positions and the sine of refraction angles $(\phi,\sin\chi)$ with
pump efficiency denotative intensity. For comparison, the phase space
presentation of emission property of the same cavity is obtained by
ray simulation {[}Fig. \ref{fig:projection}(a){]}. Those two distributions
in phase space show good consistence. Thus, the excitation is verified
numerically as inverse process of emission.

\begin{figure*}
\begin{centering}
\includegraphics[width=14cm]{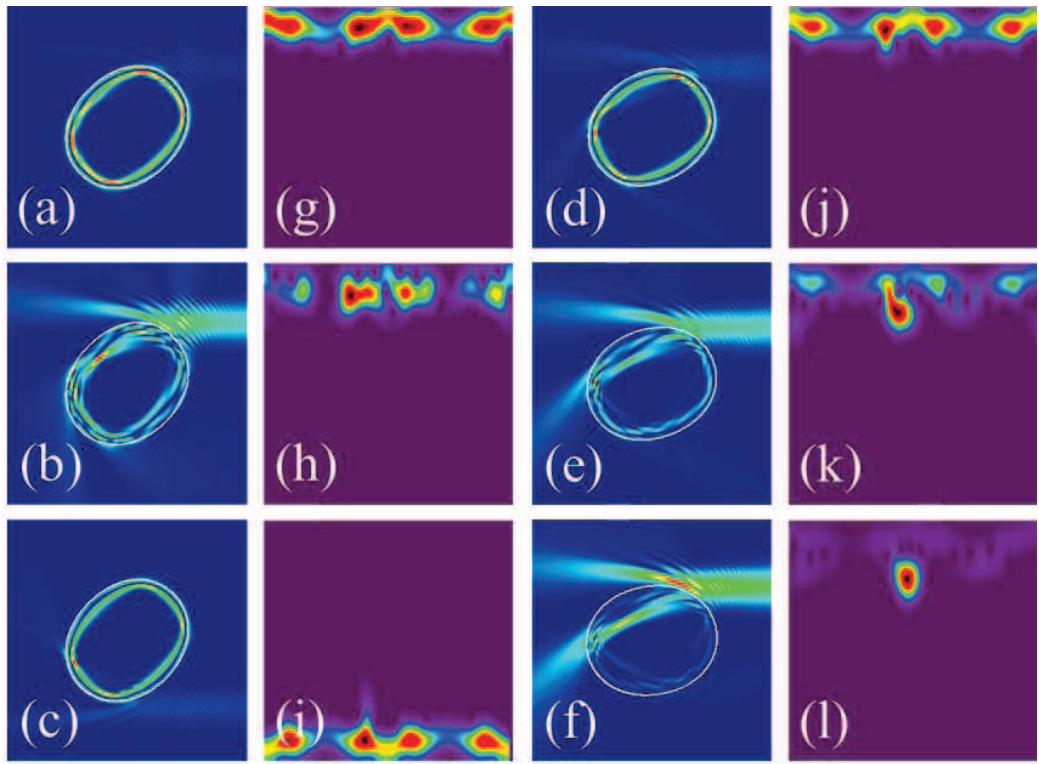}
\par\end{centering}

\caption{False color intensity patterns of excited cavity ((a), (b), (c), (d),
(e), and (f)) and corresponding Husimi projection ((g), (h), (i),
(j), (k), and (l)). The aim distances and the rotation angles are
labeled with corresponding letters in $\varepsilon=0.1$ panel of
Fig. \ref{fig:Pumpefficiency}. The coordinates of real space patterns
and phase space projection are the same with Fig \ref{fig.onoffpattern}(a)
and Fig. \ref{fig.sos}(c), respectively.\label{fig:Husimi}}
\end{figure*}

To investigate the detail of the excitation process and get a deep
physical insight of the free space coupling, Husimi projections of
the excitation (purple background picture in Fig. \ref{fig:Husimi})
are calculated for different pump condition. As demonstrated in Fig.
\ref{fig:Husimi}(g), Husimi of the optimum excitation is consistent
with the Husimi of order-2 mode [see Fig. \ref{fig.sos}(c) green
dots] in CCW half phase plane. It shows that the incident energy is
coupled into the cavity and forms resonant traveling mode {[}Fig.
\ref{fig:Husimi}(a){]}. The excitation process is also demonstrated
in the case of detuning light ($kR=50$) with the same geometry {[}Fig.
\ref{fig:Husimi}(b){]}. The Husimi projection of the detuning excitation
has four dense points which have good superposition area with the
resonance islands. Next, the excitation field corresponding to the
second maximum MII configuration, i.e. the optimal excitation on the
other side, is obtained {[}Fig. \ref{fig:Husimi}(c){]}. It is a resonant
order-2 CW mode as is shown in SOS {[}Fig. \ref{fig.sos}(c){]}. The
right side patterns in Fig. \ref{fig:Husimi} are the distributions
of excitation CCW mode by tangential incident beam at the rotating
angle of $30^{\circ}$ {[}Figs. \ref{fig:Husimi}(d), (j){]}, $20^{\circ}$
{[}Figs. \ref{fig:Husimi}(e), (k){]} and $10^{\circ}$ {[}Figs. \ref{fig:Husimi}(f),
(l){]}, respectively. The corresponding Husimi projections reflect
the change of coupling strengthens between the incident beam and the
resonance mode from strong to weak. The mode built in $30^{\circ}$
case is merely strong, while the mode of $20^{\circ}$ only has four
weak points in the SOS. As for the $10^{\circ}$, only very subtle
mode energy emerges in the SOS. As we can see in the phase space,
the more overlap between the incident beam and the resonant mode there
is, the higher excitation MII will be established. The overlap owes
to the wave character of the beam. Despite the penetration into islands
is forbidden for the refraction rays from the outside, the dynamical
tunneling effect allows it.

\section{Conclusion}

We have gotten the inner and outer field of the excitation cavity
by BEM. Scattering energy flow was also calculated with the Poynting
vector. The transmission spectrum of a certain direction can have
asymmetric lineshape because of the interference. MII distribution
agrees well with the emission picture in reverse manner in phase space.
MII distributions of cavity with different deformations also exhibit
two main excitation processes, i.e. barrier tunneling and dynamic
tunneling. Besides the frequency matching, the sufficient overlap
between the incident beam and the resonance mode is the key factor
for efficient excitation.
\begin{acknowledgments}
This work was supported by the National Natural Science Foundation
of China under Grant No.11204196/11004184, the National Fundamental
Research Program of China under Grant No. 2011CB921200, the Knowledge
Innovation Project of Chinese Academy of Sciences, the Fundamental
Research Funds for the Central Universities, and the Foundation of
He’nan Educational Committee of China under Grant No. 2011A140021.

\end{acknowledgments}

\end{document}